\documentclass[ twocolumn,
		        aps,
		        superscriptaddress,
		        pra
		       ]{revtex4-1}
		
\usepackage[colorlinks=true,urlcolor=blue,citecolor=blue,linkcolor=red]{hyperref}
\usepackage{graphicx}
\usepackage{wrapfig}
\usepackage{booktabs}
\usepackage{amsmath}
\usepackage{amssymb}
\usepackage{braket}
\usepackage[dvipsnames]{xcolor}
\usepackage[normalem]{ulem}

\usepackage{bbold}

\begin{document}
\title{Peaceful coexistence of thermal equilibrium and the emergence of time}

\author{Tommaso Favalli}\email{favalli@lens.unifi.it}
\affiliation{QSTAR, INO-CNR, and LENS, Largo Enrico Fermi 2, I-50125 Firenze, Italy}
\affiliation{Universit\`a degli Studi di Napoli \lq\lq Federico II\rq\rq, Via Cinthia 21, I-80126 Napoli, Italy}
\author{Augusto Smerzi}\email{augusto.smerzi@ino.it}
\affiliation{QSTAR, INO-CNR, and LENS, Largo Enrico Fermi 2, I-50125 Firenze, Italy}

\begin{abstract}
We consider a quantum Universe composed by a small system $S$ and
a large environment. 
It has been demonstrated that, for the vast majority of randomly chosen wave-functions of the Universe satisfying a total energy constraint,
the reduced density matrix of 
the system $S$ is given by the canonical statistical distribution. 
Here, through the Page and Wootters mechanism, we show that time and non-equilibrium dynamics  
can emerge as a consequence of the entanglement 
between the system and the environment present in the (randomly chosen) global wave-function of the Universe.
The clock is provided by the environment, which ticks the 
temporal evolution of $S$. The paradox of the peaceful coexistence of
statistical equilibrium and non-equilibrium dynamics is solved by identifying the trace over 
the environment degrees of freedom with the temporal trace over the entire history of the system $S$.
\end{abstract}

\maketitle

	\section{Introduction}
\label{Introduction}

We consider a global quantum system, the ``Universe", composed by a small system $S$ 
and a large environment $E$. It has been recently demonstrated \cite{canonical,popescu1} (see also \cite{zurek})
that, under a suitable constraint on the total energy and for the overwhelming majority of randomly chosen
pure wave-functions of the Universe, the reduced density matrix of $S$ is indistinguishable 
from a thermal canonical distribution,
a property named in \cite{canonical} as \textit{Canonical Typicality}.

Back in 1983 Don N. Page and William K. Wootters (PaW)
suggested that time can be an emergent property of entanglement between subsystems in a globally static Universe \cite{pagewootters,wootters}, a proposal that has recently attracted much attention \cite{lloydmaccone,esp1,esp2,vedral,vedraltemperature,macconeoptempoarrivo,wigner,interacting,simile,simile2,leonmaccone,review,nostro,nuovo,timedilation,scalarparticles,dirac,foti,asimmetry} as a viable route for a new description of space-time, including a new perspective 
for merging quantum clocks and gravity \cite{brukner,giacomini} (see Appendix A for a brief summary of PaW theory).

The goal of this work is to show that canonical thermal equilibrium and dynamics can coexist in a quantum Universe, with the
environment providing the clock for
the evolution of the system $S$. The dynamics is governed by the Schr\"odinger equation corrected by a non-local term which
vanishes in the limit of fixed total energy of the Universe.
The paradox of the coexistence of thermal equilibrium and non-trivial evolution of $S$ 
is solved because the trace over the degrees of freedom of the environment 
coincides with a temporal average over the entire life of $S$. The temporal dynamics of the system $S$ emerges by considering the relative states of $S$ (in Everett sense \cite{everett}) with respect to the states of the environment.

\section{Canonical Equilibrium Distribution}
\label{canonical}

Our quantum Universe 
is composed by a small system $S$ weakly interacting with a large environment $E$ 
and it is governed by the Hamiltonian 
\begin{equation}
\hat{H} = \hat{H}_E + \hat{H}_S
\end{equation}  
where $\hat{H}_E$ and $\hat{H}_S$ are the Hamiltonians of the subsystems $E$ and $S$ respectively, having dimensions $d_E \gg d_S$.
%
The relatively small interaction
between the environment and the system $S$ is neglected \cite{canonical,popescu1}.
In analogy with the standard derivation of the canonical distribution of a subsystem given a global microcanonical
distribution, we impose the total energy in the interval $\left[E,E+\delta\right]$, 
where $\delta \ll E, \Delta E^{(S)}$ (being $\Delta E^{(S)}$ the typical spacing between energy levels of the system $S$) but large enough to contain many eigenvalues of $\hat{H}_E$. This constraint corresponds to the choice of $\mathcal{H}_U \subseteq \mathcal{H}_E\otimes \mathcal{H}_S$. We consider, following \cite{canonical}, a Universe in a pure state $\ket{\Psi} \in \mathcal{H}_U$. The state of the system $S$ is obtained after tracing out the environment degrees of freedom: 
\begin{equation}
\hat{\rho}_S = Tr_E\left[\ket{\Psi}\bra{\Psi}\right].
\end{equation}
According to \cite{canonical,popescu1}, for almost every pure state $\ket{\Psi} \in \mathcal{H}_U$, the state of $S$ is well approximated by the canonical distribution
\begin{equation}\label{dadimostrare}
\hat{\rho}_S \approx
Tr_E \left[ \hat{\Omega}_U \right] \approx \frac{1}{Z} e^{-\beta \hat{H}_S}
\end{equation}
where $\beta = dS(E)/dE$ is the inverse temperature with
$S(E)$ the environment entropy, $Z = Tr \left[ e^{-\beta \hat{H}_S}\right]$ and $\hat{\Omega}_U = d_U^{-1} \hat{P}_U$ is 
the equiprobable mixed state in $\mathcal{H}_U$ with $d_U$ the dimension of the space $\mathcal{H}_U$ and $\hat{P}_U$ the projection on $\mathcal{H}_U$. Notice that equal probabilities (and random phases) are assigned in this case to all the states within $\hat{\Omega}_U$ which is thus maximally mixed in $\mathcal{H}_U$ \cite{cantyp}. Equation (\ref{dadimostrare}) implies that 
the thermal state of the small subsystem $S$ can be derived from a (randomly chosen) pure state $\ket{\Psi} \in \mathcal{H}_U$ or from 
the maximally mixed state $\hat{\Omega}_U$.

\section{Environment as a Clock}
\label{clockasenvironment}
\subsection{General framework}

We are now going to merge Canonical Typicality and PaW theory. The key point is to recognize the environment as a clock:
\begin{equation}
\hat{H}_E \equiv \hat{H}_C.
\end{equation}
Notice that in the PaW framework a good clock has to have a Hilbert space dimension larger than the dimension of the system $S$, otherwise it would no longer be possible to relate each energy eigenstate of the system to an energy eigenstate of the clock (see Appendix A and \cite{nostro}). Furthermore a good clock has to interact weakly with the system $S$ or, in the ideal case, it should not interact at all. These conditions coincide with those required for the environment by Canonical Typicality.

In the original PaW theory the global state of the Universe is an eigenstate of the total Hamiltonian with zero eigenvalue, i.e. $\hat{H}\ket{\Psi}=0$. 
In this work we follow a slightly different path and we weakly relax the PaW constraint considering the total energy of the Universe
within the energy shell $\left[E,E+\delta\right]$ where $\delta \ll E,\Delta E^{(S)}$ but large enough to contain many energy eigenvalues of the clock $\equiv$ environment $C$. In this framework we find a non-local Schrödinger-like evolution for the relative state of $S$ that reduces to the usual Schrödinger dynamics for times $t-t_0 \ll 1/\delta$ (where $t_0$ is the initial time) or for all times in the limit $\delta \rightarrow 0$.

We define the time states in the Hilbert space of the clock $\equiv$ environment using the approach developed in \cite{nostro,pegg}.
We assume that the Hamiltonian of the environment has non-degenerate energy eigenstates with rational energy ratios
\begin{equation}\label{numraz}
\frac{E^{(C)}_i }{E^{(C)}_1} = \frac{A_i}{D_i},
\end{equation}
where $A_i$ and $D_i$ are integers with no common factors and $E^{(C)}_0 =0$. This implies that all energy values are integer multiples of a (arbitrarily small) step ($\hslash=1$):
\begin{equation}\label{ei}
E^{(C)}_i = r_i \frac{2\pi}{T}
\end{equation}
where $T=\frac{2\pi r_1}{E_1}$, $r_i = r_1\frac{A_i}{D_i}$ for $i>1$ (with $r_0=0$) and $r_1$ is 
equal to the lowest common multiple of the values of $D_i$. In this space we define the states 
\begin{equation}
\ket{t_m}  = \frac{1}{\sqrt{p+1}}\sum_{i=0}^{p}e^{-i E^{(C)}_i t_m}\ket{E^{(C)}_i}
\end{equation}
with $p+1 = d_{C}$,
$t_m = t_0 + mT /(s+1)$, 
$m=0,1,2,...,s$ and $s+1 \ge r_p$. The number of states $\ket{t_m}$ is therefore greater than the number of energy states 
in $\mathcal{H}_{C}$ and the $s+1$ values of $t_m$ are uniformly distributed over $T$. These states are not orthogonal but provide an overcomplete basis in $C$ with the resolution of the identity
\begin{equation}\label{pomidentity}
\frac{p+1}{s+1} \sum_{m=0}^{s} \ket{t_{m}}\bra{t_{m}} = \mathbb{1}_{C}.
\end{equation}
In order to obtain a continuous flow of time we can now consider the limit $s \rightarrow \infty $ and define
\begin{equation}\label{alphastateinf}
\ket{{t}} = \sum_{i=0}^{p} e^{- i E^{(C)}_i t}\ket{E^{(C)}_i}
\end{equation}
where $t$ can now take any real value from $t_0$ to $t_0 + T $. In this limit the resolution of the identity (\ref{pomidentity}) becomes
\begin{equation}\label{newresolution}
\frac{1}{T} \int_{t_0}^{t_0+T} d t \ket{t} \bra{t} = \mathbb{1}_{C} 
\end{equation}
and the states $\ket{t}$ provide again an overcomplete basis with $\braket{t|t'}=\sum_{n}e^{i E^{(C)}_n (t - t')}$.
We notice that, if we would consider non-rational ratios of energy levels, the resolutions of the identity (\ref{pomidentity}) and (\ref{newresolution}) are no longer exact and the time states $\ket{t_m}$ and $\ket{t}$ do not provide an overcomplete basis in $C$. However, 
since any real number can be approximated with arbitrary precision by a ratio between two rational numbers, the residual terms in the resolutions of the identity can be arbitrarily reduced \cite{nostro}.

\subsection{Random Universe and dynamics}

Here we show that a Universe in a (randomly chosen) pure state is compatible with the emergence of time and a non-trivial 
dynamical evolution of the system $S$. 
The global state of the Universe is
\begin{equation}\label{statoglobale}
\ket{\Psi} = \sum_{j}\sum_{i} c_{ij} \ket{E^{(C)}_i} \ket{E^{(S)}_j}
\end{equation}
where we take the coefficients $c_{ij}$ distributed as in \cite{canonical}.
With $\ket{\Psi}=\ket{\Phi}/||\ket{\Phi}||$ and $\ket{\Phi} =\sum_{j}\sum_{i} \tilde{c}_{ij} \ket{E^{(C)}_i} \ket{E^{(S)}_j}$,
the real and imaginary parts of the coefficients $\tilde{c}_{ij}=c_{ij}||\ket{\Phi}||$ are chosen as
independent real Gaussian random variables with mean zero and variance $1/2$ for the 
values of $i, j$ such that
$E^{(C)}_i + E^{(S)}_j \in \left[E,E+\delta\right]$ and $\tilde{c}_{ij}=0$ otherwise. 
Since
$\delta  \ll \Delta E^{(S)}_j = E^{(S)}_{j+1} - E^{(S)}_{j}~ \forall j$
and considering that the spectrum of $\hat{H}_{C}$ is much denser than the spectrum of $\hat{H}_{S}$, 
the constraint on the total energy implies that each level of the system is coupled with several neighbour levels of the clock. 
The random choice of the coefficients provides Canonical Typicality \cite{canonical}: here we show that is also sufficient
to provide the temporal dynamics for the relative state (in Everett sense \cite{everett}) of the system $S$.

The action of the global Hamiltonian $\hat{H}$ on the global state $\ket{\Psi}$ gives
\begin{equation}\label{applicazionehtot}
\begin{split}
\hat{H}\ket{\Psi} & = 
\left( \hat{H}_{C} + \hat{H}_S  \right)\sum_{j}\sum_{i \in I_j} c_{ij} \ket{E^{(C)}_i} \ket{E^{(S)}_j} =
\\&= \sum_{j}\sum_{i \in I_j} c_{ij} \left( E^{(C)}_i + E^{(S)}_j  \right) \ket{E^{(C)}_i} \ket{E^{(S)}_j} =
\\&= E \ket{\Psi} + \sum_{j}\sum_{i \in I_j} c_{ij} \Delta_{ij} \ket{E^{(C)}_i} \ket{E^{(S)}_j}
\end{split} 
\end{equation} 
where $I_j$ is the set of the environment levels such that $E^{(C)}_i \in \left[E- E^{(S)}_j , E- E^{(S)}_j + \delta \right]$ 
and where we have written $E^{(C)}_i + E^{(S)}_j  = E + \Delta_{ij}$ with $\Delta_{ij} \in \left[0,\delta\right]$. The relative state of the system $S$ at a certain ``time" $t$ is defined by
\begin{equation}\label{defstatorelativo}
\ket{\phi (t)}_S = \braket{t|\Psi}
\end{equation}
(notice that (\ref{defstatorelativo}) is still a pure state) and its time evolution can be easily calculated:

\begin{widetext}
\begin{equation}\label{eqschroedinger}
\begin{split}
i \frac{\partial}{\partial t} \ket{\phi(t) }_S &= \frac{\partial}{\partial t} \sum_{k}^{d_{C}}\bra{E_k}e^{iE_k t}\ket{\Psi} = - \sum_{k}^{d_{C}}\bra{E_k}E_ke^{i E_k t}\ket{\Psi}= \\&
=-\bra{t}\hat{H}_{C}\ket{\Psi}=
\hat{H}_S \braket{t|\Psi} - \bra{t}\hat{H}\ket{\Psi}  =\\ \\&
= \left(\hat{H}_S - E\right) \ket{\phi(t)}_S - \bra{t}\sum_{j}\sum_{i \in I_j} c_{ij} \Delta_{ij} \ket{E^{(C)}_i} \ket{E^{(S)}_j} 
\end{split}
\end{equation}
where we have used (\ref{alphastateinf}), $\hat{H} = \hat{H}_{C} + \hat{H}_S$, (\ref{defstatorelativo}) and (\ref{applicazionehtot}). By defining the operator $\hat{\Delta} =  \sum_{j} \sum_{i \in I_j} \Delta_{ij} \ket{E^{(C)}_i} \ket{E^{(S)}_j} \bra{E^{(C)}_i}  \bra{E^{(S)}_j}$
and removing the term related to $E$ which gives an irrelevant phase factor in the evolution of $S$, (\ref{eqschroedinger}) becomes the time non-local Schrödinger-like equation:

\begin{equation}\label{eqschrodinger2}
i \frac{\partial}{\partial t} \ket{\phi(t) }_S =  \hat{H}_S  \ket{\phi(t)}_S - \frac{1}{T} \int_{t_0}^{t_0+T} dt' \hat{\Delta}(t,t')\ket{\phi(t')}_S
\end{equation}
where $\hat{\Delta}(t,t') = \bra{t}\hat{\Delta}\ket{t'}$. 
The last term in the right-hand side of the equation is an integral operator acting on $S$. 


For times $t-t_0 \ll 1/\delta$ (and so $t-t_0 \ll 1/\Delta_{ij}$ for typical $\Delta_{ij}$) 
(\ref{eqschrodinger2}) reduces to the ordinary Schr\"odinger equation. Indeed we have:
\begin{equation}\label{17new}
\begin{split}
\ket{\phi(t)}_S &= \braket{t|\Psi} = \bra{t_0}e^{i\hat{H}_C(t-t_0)}\ket{\Psi}= e^{-i\hat{H}_S(t-t_0)}\bra{t_0}e^{i\hat{H}(t-t_0)}\ket{\Psi} \\&
= e^{-i\hat{H}_S(t-t_0)}\bra{t_0}\sum_{j}\sum_{i \in I_j}c_{ij} e^{i(E^{(C)}_i + E^{(C)}_j)(t-t_0)}\ket{E^{(C)}_i}\ket{E^{(S)}_j}=\\&
= e^{-i\hat{H}_S(t-t_0)}\bra{t_0}\sum_{j}\sum_{i \in I_j}c_{ij} e^{iE(t-t_0)} e^{i\Delta_{ij}(t-t_0)} \ket{E^{(C)}_i}\ket{E^{(S)}_j}
\end{split}
\end{equation}
\end{widetext}
where we used (\ref{defstatorelativo}), $\hat{H} = \hat{H}_{C} + \hat{H}_S$ and $E^{(C)}_i + E^{(S)}_j  = E + \Delta_{ij}$. For $t-t_0 \ll 1/\delta$, considering $e^{i\Delta_{ij}(t-t_0)}\simeq 1$ and removing the irrelevant global phase factor $e^{iE(t-t_0)}$, (\ref{17new}) becomes

\begin{equation}
\ket{\phi(t)}_S \simeq e^{-i \hat{H}_S(t-t_0)}\ket{\phi(t_0)}_S 
\end{equation}
which provides the Schr\"odinger evolution for the system $S$. In Section III.D we'll briefly discuss the effect of the non-local term in equation (\ref{eqschrodinger2}) for times $t- t_0 \ge 1/\delta$.

Equation (\ref{eqschrodinger2}) can be explicitly solved: with $E^{(C)}_i + E^{(S)}_j = \Delta_{ij}$ (the term related to $E$ has been removed) we obtain (see Appendix B):

\begin{equation}\label{metodo2}
\begin{split}
\ket{\phi (t)}_S =  \sum_{j} \alpha_j(t)  e^{-i  E^{(S)}_j t}  \ket{E^{(S)}_j}
\end{split}
\end{equation}
with
\begin{equation}\label{alphat}
\alpha_j (t) = \sum_{i \in I_j} c_{ij}e^{i \Delta_{ij} t} .
\end{equation}
Equation (\ref{metodo2}) provides an additional time dependence to the Schr\"odinger evolution through the coefficients $\alpha_j(t)$. 
In the case $t-t_0 \ll 1/\delta$ we have
$\alpha(t) \simeq  \sum_{i \in I_j}c_{ij}e^{i\Delta_{ij}t_0} \equiv \alpha_j(t_0)$ and the state (\ref{metodo2}) becomes $ \ket{\phi (t)}_S \simeq \sum_{j} \alpha_j(t_0)  e^{-i E^{(S)}_j t} \ket{E^{(S)}_j}$ where we recognize again the Schr\"odinger evolution for the system $S$.

We emphasize that (\ref{eqschrodinger2}) does not preserve the norm of the state $\ket{\phi(t)}_S$ over time. Indeed, calculating the scalar product $\braket{\phi(t)|\phi(t)}$ through (\ref{metodo2}) we obtain
\begin{widetext}
\begin{equation}\label{evoluzionenorma}
\begin{split}
\braket{\phi(t)|\phi(t)} &= \sum_{j} \left|\alpha_j(t)\right|^2 = \sum_{j} \sum_{i \in I_j} \sum_{k \in I_j} c_{ij}c^{*}_{kj} e^{i(\Delta_{ij} - \Delta_{kj})t} =\\&
=  \sum_{j} \sum_{i \in I_j} \sum_{k \in I_j}|c_{ij}||c_{kj}|\cos((\Delta_{ij} - \Delta_{kj})t - \Delta\varphi^{(j)}_{ik})
\end{split} 
\end{equation}
\end{widetext}
where $c_{ij}=|c_{ij}|e^{i\varphi_{ij}}$, $c_{kj}=|c_{kj}|e^{i\varphi_{kj}}$ and $\Delta\varphi^{(j)}_{ik} = \varphi_{kj} - \varphi_{ij}$. However, the corrections remain small when $t-t_0 \ll 1/\delta$ (and vanish in the limit $\delta\rightarrow 0$) 
being	
$\braket{\phi(t)|\phi(t)} \simeq \sum_{j} \sum_{i,k \in I_j} c_{ij}c^{*}_{kj}e^{i(\Delta_{ij} - \Delta_{kj})t_0}=\sum_{j} |\alpha_j(t_0)|^2$ that is different from the unity but (approximately) constant over time. We notice that a similar problem arose, for different reasons, in \cite{interacting} where it was 
handled by introducing a new definition for the inner product. 

To summarize, the environment can provide the clock for the dynamical evolution of the system $S$. 
The state of $S$ conditioned to a certain value of the clock through (\ref{defstatorelativo}) consists of a pure state obeying a non-local
dynamical Schrödinger-like equation (that reduces to the standard Schrödinger equation for $t-t_0 \ll 1/\delta$). Nevertheless, after tracing out the degrees of freedom of the clock $\equiv$ environment, we find the system in a state of thermal equilibrium provided by the canonical distribution. This compatibility is simply explained by the fact that 
the trace over the environment degrees of freedom is equivalent to the trace over all times. Indeed we have (see Appendix D):
\begin{equation}\label{mediatemporale}
\hat{\rho}_S = Tr_C\left[\ket{\Psi}\bra{\Psi}\right] = \frac{1}{T} \int_{t_0}^{t_0+T} dt \braket{t|\Psi}\braket{\Psi|t}   \approx \frac{1}{Z} e^{-\beta \hat{H}_S}
\end{equation}
where again $\beta = dS(E)/dE$ is the inverse temperature with $S(E)$ the clock $\equiv$ environment entropy and $Z = Tr \left[ e^{-\beta \hat{H}_S}\right]$ as in equation (\ref{dadimostrare}). 

\subsection{Initial conditions for $S$}

The merging of Canonical Typicality and PaW imposes a constraint on the allowed initial conditions of the 
subsystem $S$. 
The initial condition for the state (\ref{metodo2}) is 
\begin{equation}\label{defstatoiniziale}
\ket{\phi (t_0)}_S =   \sum_{j}\alpha_j(t_0) e^{-iE^{(S)}_jt_0 } \ket{E^{(S)}_j}
\end{equation}
with
\begin{equation}\label{alpha}
\alpha_j(t_0) = \sum_{i \in I_j} c_{ij} e^{i\Delta_{ij}t_0} .
\end{equation}
The reduced density matrix $\hat{\rho}_S$ of the subsystem $S$
is
\begin{equation}\label{mdr}
\begin{split}
\hat{\rho}_S = Tr_C\left[\ket{\Psi}\bra{\Psi}\right] 
= \sum_{j} \sum_{i\in I_j}\left|c_{ij} \right|^2 \ket{E^{(S)}_j}\bra{E^{(S)}_j}
\end{split}
\end{equation}
where, following \cite{canonical}, we have considered that the relevant energy intervals for the clock $\equiv$ environment coupled with each level $E^{(S)}_j$ of $S$ are pairwise disjoint (which is a consequence of $\delta  \ll \Delta E^{(S)}_j = E^{(S)}_{j+1} - E^{(S)}_{j}~ \forall j$). Canonical Typicality implies:
\begin{equation}\label{condcoeff}
\sum_{i \in I_j}\left|  c_{ij} \right|^2 \approx \frac{1}{Z} e^{-\beta E^{(S)}_j}
\end{equation}
which constrains the initial conditions by selecting a set of allowed $\alpha_j(t_0)$ through (\ref{alpha}). It is crucial to notice that 
the condition (\ref{condcoeff}) contrains the sum of the absolute values of the coefficients $c_{ij}$ and therefore leaves a large margin of freedom 
on the possible values of $\alpha_j(t_0)$. 	
In conclusion, Canonical Typicality states that the reduced density matrix $\hat{\rho}_S=Tr_{C}\left[\ket{\Psi}\bra{\Psi}\right]$ of the \textit{overwhelming majority} of the pure wave-functions $\ket{\Psi} \in \mathcal{H}_U \subseteq \mathcal{H}_C\otimes\mathcal{H}_S$ is canonical. This means that the overwhelming majority of the randomly chosen coefficients $c_{ij}$ satisfies the condition (\ref{condcoeff}) which, in our framework, is
consistent with a non-trivial dynamical evolution of the system $S$. 

\subsection{A toy model}

We look now at a simple example assuming that the subsystem $S$ consists of a one-dimensional harmonic oscillator 
with Hamiltonian $\hat{H}_S = \frac{\hat{P}^2}{2m} + \frac{1}{2}m\omega^2\hat{X}^2$. 
We assume that the dynamics is confined among the two lowest energy levels of the oscillator and therefore the global state of the Universe (\ref{statoglobale}), satisfying the constraint on the total energy, is $\ket{\Psi} = \sum_{i \in I_0} c_{i0} \ket{E^{(C)}_i}\ket{0^{(S)}} + \sum_{i \in I_1} c_{i1} \ket{E^{(C)}_i}\ket{1^{(S)}}$.

With (\ref{defstatoiniziale}) and $t_0 = 0$, we look at the (pure) initial state of the subsystem $S$: $\ket{\phi(0)}_S = \alpha_0(0) \ket{0^{(S)}} + \alpha_1(0) \ket{1^{(S)}}$, 	
where we set the initial values $\alpha_0(0)$ and $\alpha_1(0)$ according to (\ref{alpha}). Thanks to (\ref{metodo2}) and (\ref{alphat}), for the conditioned state of $S$ at a generic time $t$, we have

\begin{widetext}
\begin{equation}\label{statotoymodel}
\begin{split}
\ket{\phi(t)}_S &= \alpha_0(t) e^{-i E^{(S)}_0t} \ket{0^{(S)}} +  \alpha_1(t) e^{-i E^{(S)}_1 t} \ket{1^{(S)}} =\\&
= \sum_{i \in I_0} c_{i0} e^{-i (E^{(S)}_0 - \Delta_{i0})t} \ket{0^{(S)}} + \sum_{i \in I_1} c_{i1} e^{-i (E^{(S)}_1 - \Delta_{i1})t} \ket{1^{(S)}}.
\end{split}
\end{equation}
We recall here that the state (\ref{statotoymodel}) is not normalized. To restore the normalization (for $t\ll 1/\delta$ where the norm is approximately preserved) we should divide (\ref{statotoymodel}) by $\sqrt{\alpha^2_0(0) + \alpha^2_1(0)}$. However, for the sake of simplicity, we proceed  with the calculation without considering the normalization.

We look now at the time dependence of the expectation value of the position operator
$\langle \hat{X} \rangle_{t} = \bra{\phi(t)} \hat{X} \ket{\phi(t)}$ and obtain (see Appendix E):

\begin{equation}\label{evestgrande}
\begin{split}
\langle \hat{X} \rangle_{t} &=\sqrt{\frac{2}{m\omega}} |\alpha_{0}(t)||\alpha_{1}(t)|\cos(\omega t - \Delta\phi(t)) =\\&
=  \sqrt{\frac{2}{m\omega}} \sum_{i \in I_0} \sum_{k \in I_1}  |c_{i0}||c_{k1}|\cos((\omega  + (\Delta_{i0} - \Delta_{k1}))t - \Delta\varphi^{(0,1)}_{ik})
\end{split}
\end{equation}
where we have considered $\alpha_{0}(t)= |\alpha_{0}(t)|e^{i\phi_{0}(t)}$, $\alpha_{1}(t)= |\alpha_{1}(t)|e^{i\phi_{1}(t)}$, $\Delta\phi (t)= \phi_{1}(t) - \phi_{0}(t)$, $c_{i0}= |c_{i0}|e^{i\varphi_{i0}}$, $c_{k1}= |c_{k1}|e^{i\varphi_{k1}}$, $\Delta\varphi^{(0,1)}_{ik} = \varphi_{k1} - \varphi_{i0}$ and $E^{(S)}_1 - E^{(S)}_0 = \omega$. 
For times $t \ll 1/|\Delta_{i0} - \Delta_{k1}|$, up to first order of approximation in $t( \Delta_{i0} - \Delta_{k1})$, (\ref{evestgrande}) reduces to (see Appendix F):

\begin{multline}\label{dimappe}
\langle \hat{X} \rangle_{t} \simeq  \sqrt{\frac{2}{m\omega}}  |\alpha_0(0)||\alpha_1(0)|\cos(\omega t - \Delta \phi(0)) -  \sqrt{\frac{2}{m\omega}} \sum_{i \in I_0} \sum_{k \in I_1}  |c_{i0}||c_{k1}|t( \Delta_{i0} - \Delta_{k1} )\sin(\omega t  - \Delta\varphi^{(0,1)}_{ik})  
\end{multline}
\end{widetext}
where $\alpha_0(0) = |\alpha_0(0)|e^{i \phi_0(0)}$, $\alpha_1(0) = |\alpha_1(0)|e^{i \phi_1(0)}$, $\Delta \phi(0) = \phi_1(0) - \phi_0(0)$ and where we used again (\ref{alpha}). Equation (\ref{dimappe}) indicates, as expected, that the expectation value $\langle \hat{X} \rangle_{t}$ oscillates between $\pm \sqrt{\frac{2}{m\omega}} |\alpha_0(0)||\alpha_1(0)|$ with frequency $\omega$ (apart from small corrections) and this is not surprising since we know that for $t\ll 1/\delta$ the system $S$ exhibits a Scr\"odinger-like evolution. 

If we trace over the clock $\equiv$ environment degrees of freedom, which corresponds to a time average, 
for the overwhelming majority of the randomly chosen coefficients $c_{ij}$ 
we obtain, thanks to (\ref{mediatemporale}), (\ref{mdr}) and (\ref{condcoeff}), the canonical mixed density matrix for the subsystem $S$: 
\begin{equation}\label{ultimaes}
\begin{split}
\hat{\rho}_S & = \frac{1}{T} \int_{0}^{T} dt \braket{t|\Psi}\braket{\Psi|t} = \\&=  \sum_{i \in I_0}|c_{i0}|^2 \ket{0^{(S)}}\bra{0^{(S)}} + \sum_{i \in I_1}|c_{i1}|^2 \ket{1^{(S)}}\bra{1^{(S)}} \approx  \\&
\approx \frac{1}{Z}\left( e^{-\beta E^{(S)}_0} \ket{0^{(S)}}\bra{0^{(S)}} + e^{-\beta E^{(S)}_1} \ket{1^{(S)}}\bra{1^{(S)}} \right) .
\end{split}
\end{equation}

\section{Discussion}

\subsection{On conditional probabilities}
An important point in the PaW mechanism concerns conditional probabilities. In the original PaW proposal the probability to obtain the outcome $a$ when measuring the observable $\hat{A}$ (with $\hat{A}\ket{a}=a\ket{a}$) on the subspace $S$ at a certain clock time $t$ is: 
\begin{equation}
\begin{split}
p(a \: on \: S \: | \: t \: on \: C) = \frac{p(a \: on \: S, \: t \: on \: C)}{p( t \: on \: C)}
\end{split}
\end{equation}
that is the conditional probability of obtaining $a$ on $S$ given that the clock $C$ shows $t$. 

This aspect of the theory has been criticised by K. V. Kuchar \cite{kuchar} who emphasized that the PaW mechanism is not able to provide the correct propagators when considering multiple measurements. Indeed measurements of the system at two times will give the wrong statistics because the first measurement \lq\lq collapses\rq\rq the time state and freezes the system. Two possible ways out of this problem have been proposed: the first in \cite{gppt} (which we call GPPT theory in the following) with an experimental illustration in \cite{esp1} and the second in \cite{lloydmaccone}. 

We focus now on the GPPT proposal. As pointed out in \cite{esp1} one of the main ingredients in the GPPT theory 
is the averaging over the abstract coordinate time (the \lq\lq external time\rq\rq) in order to eliminate any external time dependence in the observables. So, in the GPPT proposal, the probability to obtain the outcome $a$ when measuring the observable $\hat{A}$ on the subsystem $S$ conditioned to having the outcome $t$ on $C$ is given by \cite{gppt}:

\begin{equation}\label{cpGPPT}
	p(a \: on \: S \: | \: t \: on \: C) = \frac{\int d\theta \: Tr \left[ \hat{P}_{a,t}(\theta) \hat{\rho} \right]}{\int d\theta \: Tr \left[ \hat{P}_{t}(\theta) \hat{\rho} \right]}
\end{equation}
where $\theta$ is the external time, $\hat{\rho}=\ket{\Psi}\bra{\Psi}$ is the global state of the Universe, $\hat{P}_{t}(\theta)=\hat{U}^{\dagger} (\theta) \hat{P}_{t} \hat{U}(\theta)$ (with $\hat{U}(\theta) = e^{-i\hat{H}\theta}$) is the projector relative to the result $t$ for a clock measurement at external time $\theta$ and $\hat{P}_{a,t}(\theta)=\hat{U}^{\dagger} (\theta) \hat{P}_{a,t} \hat{U}(\theta)$ is the projector relative to the result $a$ for a measurement on $S$ and $t$ for a measurement on $C$ at external time $\theta$ (we are working here in the Heisenberg picture with respect to the external time $\theta$). The generalization of equation (\ref{cpGPPT}) to multiple time measurements is given by \cite{gppt}:

\begin{equation}\label{cpGPPT2}
\begin{split}
& p(a_f \: on \: S \: | \:t_f \: on \: C,a_i,t_i) =\\&
= \frac{\int d\theta \int d\theta' \: Tr \left[\hat{P}_{a_f,t_f}(\theta)  \hat{P}_{a_i,t_i}(\theta') \hat{\rho} \hat{P}_{a_i,t_i}(\theta') \right]}{\int d\theta \int d\theta' \: Tr \left[\hat{P}_{t_f}(\theta) \hat{P}_{a_i,t_i}(\theta') \hat{\rho} \hat{P}_{a_i,t_i}(\theta') \right]}
\end{split}
\end{equation}
which provides the conditional probability of obtaining $a_f$ on the system $S$ at clock time $t_f$, given that a \lq\lq previous\rq\rq joint measurement of $S$ and $C$ returns $a_i$, $t_i$.  

The GPPT proposal needs a global state $\hat{\rho}$ commuting with the global Hamiltonian and then it can be applied to our framework in the limit $\delta \rightarrow 0$. 
We finally notice and emphasize that, in our framework, the fact of having non-orthogonal time states does not constitute a problem in the application of the GPPT theory. Indeed when we calculate the relative state of the system $S$ to a certain clock value $t$ through (\ref{defstatorelativo}), we find no contributions from times $\ne t$ and interference phenomena are not present even if the time states are not orthogonal \cite{nostro}.

\subsection{Non-observable Universe as a clock}

A condition to merge Canonical Typicality and PaW time is to have an environment that is negligibly interacting with $S$.
This is certainly not the case in our everyday life where decoherence due to the
surrounding environment is often non-negligeable. A possible choice of a good clock non-interacting with $S$ is the non-observable Universe (namely, a part of the global system that is outside the light cone of $S$).
In this respect, it is intriguing to notice that the recent observations on the cosmic microwave background \cite{planck} together with the inflationary paradigm indicate that at the beginning of cosmic inflation the Universe was in a pure state with highly-correlated quantum fluctuations \cite{bianchi}. 
Furthermore, it has been suggested that the assumption that the observable and the non-observable Universe might be entangled provides an argument in support of inflation \cite{vedraltemperature}, \cite{vedralinflation}. It is therefore somehow natural to speculate that the non-observable Universe acts as a clock for the observable Universe. Indeed in this framework the two requirements for a ``good clock" are satisfied: the clock and the system $S$ are non-interacting 
and, in addition, the dimension of the clock is presumably larger than the dimension of $S$ (i.e. the the non-observable Universe is bigger than the observable Universe). 
A very simple estimate supports the consistency of the scenario.  
The spacing between the energy levels in the clock space is: 
\begin{equation}\label{spacingenergy}
\delta E^{(C)}_i = E^{(C)}_{i+1}  + E^{(C)}_i = \frac{2\pi \hslash}{T} \left( r_{i+1} - r_i \right)  
\end{equation}
where $r_{i+1} - r_i$ is an integer. Notice that $\delta E^{(C)}_{min} = 2\pi \hslash/T$ is the minimum energy step value, so that all other energy values can be considered as integer multiples of this minimum step. The $\delta E^{(C)}_{min}$ is inversely proportional to the value of $T$, that is the time taken by the clock to return to its initial state (notice that the framework we introduced in Section III leads to a cyclical flow of time).
So, considering the global system as the whole Universe, 
we can relate $T$ to the current age of the Universe $T_U$ by assuming $T \ge T_{U} \sim 13.8 \times 10^{9} y \simeq 4.35 \times 10^{17} s$ (and $t_0 = 0$ the instant of the Big Bang).
This leads to 
\begin{equation}\label{33}
\delta E^{(C)}_{min} = \frac{2\pi \hslash}{T} \le \frac{2\pi \hslash}{T_{Universe}} \simeq 1.5 \times 10^{-51} J .
\end{equation}
This upper limit for $\delta E^{(C)}_{min}$ is very small compared to other energies on atomic scale and fits into our framework of constructing the spectrum of the Hamiltonian $\hat{H}_{C}$ with integer multiples of a minimum energy step. 



\section{Conclusions}
\label{Conclusions}

In this work we have merged Canonical Typicality and the PaW quantum time.
We consider a quantum Universe made by a small system $S$ and a large environment which serves as a clock for $S$. Thanks to Canonical Typicality we know that for almost all pure states in which the whole Universe can be, after tracing over the environment the system $S$ is in a state of equilibrium described by the canonical distribution. In the same scenario we find a Schrödinger-like evolution corrected by a non-local term for the relative state of $S$ with respect to the clock $\equiv$ environment. 
Canonical Typicality and dynamics can coexist because in our protocol 
the action of tracing out the environment is equivalent to tracing over all times: the trace over the environment coincides with a temporal average.

\section*{Acknowledgements}
    We acknowledge funding from the H2020 QuantERA ERA-NET Cofund in Quantum Technologies projects QCLOCKS.

\appendix

	\section{Summary of Page and Wootters theory}
\label{PaW}

We give here a brief review of the PaW theory. We consider the whole Universe as being in a stationary state with zero eigenvalue (and therefore there is no need for an external time), consistently with the the Wheeler-DeWitt equation
\begin{equation}\label{wdw}
\hat{H}\ket{\Psi} = 0 
\end{equation}
where $\hat{H}$ and $\ket{\Psi}$ are the Hamiltonian and the state of the Universe respectively.
We can then divide the Universe into two non-interacting subsystems, the clock $C$ and the rest of the Universe $S$, and thus the total Hamiltonian can be written as
\begin{equation}\label{h+h1}
\hat{H}=\hat{H}_C + \hat{H}_S
\end{equation}  
where $\hat{H}_C$ and  $\hat{H}_S$ are the Hamiltonians acting on $C$ and $S$ respectively.
The condensed history of the system $S$ is written through the entangled global stationary state $\ket{\Psi} \in \mathcal{H} \subseteq \mathcal{H}_C \otimes \mathcal{H}_S$ (which satisfies the constraint (\ref{wdw})) as follows:
\begin{equation}\label{statoespanso}
\ket{\Psi} = \frac{1}{\sqrt{d_C}} \sum_{\tau} \ket{\tau}_C \otimes \ket{\phi_{\tau}}_S 
\end{equation}
where $d_C$ is the dimension of the clock subspace and the states $\left\{ \ket{\tau}_C \right\}$ are eigenstates of the clock observable. We notice here that a \lq\lq good clock\rq\rq must have $d_C \gg d_S$. 
Indeed, writing $\ket{\Psi}$ in the energy eigenbasis, one finds that with $d_C \le d_S$ it would not be possible to couple every energy state of $S$ to an energy state of $C$ satisfying the constraint (\ref{wdw}) and some states of $S$ would be excluded from the dynamics \cite{nostro}.

In this framework the relative state (in Everett sense \cite{everett}) of the subsystem $S$ with respect to the clock $C$ can be obtained via conditioning

\begin{equation}\label{defstatos}
\ket{\phi_{\tau}}_S = \frac{\braket{\tau|\Psi}}{1/\sqrt{d_C}} .
\end{equation}
Note that, as mentioned before, equation (\ref{defstatos}) is the Everett \textit{relative state} definition of the subsystem $S$ with respect to the clock system $C$. As pointed out in \cite{vedral}, this kind of projection has nothing to do with a measurement. Rather, $\ket{\phi_{\tau}}_S$ is a state of $S$ conditioned to the clock $C$ in the state $\ket{\tau}_C$. 
Then, from equations (\ref{wdw}), (\ref{h+h1}) and (\ref{defstatos}), it is possible to derive the Schrödinger evolution for the relative state of the subsystem $S$ with respect to the clock $C$ ($\hslash=1$):

\begin{equation}
\ket{\phi_{\tau}}_S = e^{-i\hat{H}_S (\tau -\tau_0)} \ket{\phi_{\tau_0}}_S 
\end{equation}
where $\ket{\phi_{\tau_0}}_S = \sqrt{d_C} \braket{\tau_0|\Psi}$, being $\ket{\tau_0}_C$ the clock eigenstate taken as initial time.

We want to emphasize here that, as pointed out in \cite{lloydmaccone}, in the PaW approach the zero eigenvalue of $\hat{H}$ does not play a special role in identifying the global state $\ket{\Psi}$. Indeed up to an irrelevant global phase in the dynamics of $\ket{\phi_{\tau}}_S$, the global state $\ket{\Psi}$ can be obtained also by imposing the constraint $\hat{H}\ket{\Psi} = E \ket{\Psi}$ with real $E$. This consideration is useful for the arguments we address in the main text.

\begin{widetext}
\section{Proof of equation (\ref{metodo2})}

Here we show that the state 
\begin{equation}\label{soluzione}
\ket{\phi (t)}_S =   \sum_{j} \sum_{i \in I_j} c_{ij}  e^{-i ( E^{(S)}_j - \Delta_{ij} )t}  \ket{E^{(S)}_j}
\end{equation}
is a solution for the equation 
\begin{equation}\label{eqdimezzo}
i \frac{\partial}{\partial t} \ket{\phi(t) }_S = \hat{H}_S  \ket{\phi(t)}_S - \bra{t}\hat{\Delta}\ket{\Psi} 
\end{equation}
(where $\hat{\Delta} =  \sum_{j} \sum_{i \in I_j} \Delta_{ij} \ket{E^{(C)}_i} \ket{E^{(S)}_j} \bra{E^{(C)}_i}  \bra{E^{(S)}_j}$) and for equation (\ref{eqschrodinger2}). In the first case we substitute (\ref{soluzione}) in (\ref{eqdimezzo}) thus obtaining

\begin{multline}
i \left( -i \sum_{j}\sum_{i\in I_j} c_{ij} E^{(S)}_j e^{-i(E^{(S)}_j - \Delta_{ij})t}\ket{E^{(S)}_j} + i  \sum_{j}\sum_{i\in I_j} c_{ij} \Delta_{ij} e^{-i(E^{(S)}_j - \Delta_{ij})t}\ket{E^{(S)}_j} \right) = \\= \hat{H}_S \ket{\phi (t)}_S - \bra{t}\sum_{j} \sum_{i\in I_j} c_{ij} \Delta_{ij} \ket{E^{(C)}_i} \ket{E^{(S)}_j}
\end{multline}
\begin{multline}
\Rightarrow \hat{H}_S \ket{\phi (t)}_S - \sum_{j}\sum_{i\in I_j} c_{ij} \Delta_{ij} e^{-i(E^{(S)}_j - \Delta_{ij})t}\ket{E^{(S)}_j} =  \hat{H}_S \ket{\phi (t)}_S -  \bra{t}\sum_{j} \sum_{i\in I_j} c_{ij} \Delta_{ij} \ket{E^{(C)}_i} \ket{E^{(S)}_j} .
\end{multline}
So we have:

\begin{equation}\label{44}
\sum_{j}\sum_{i\in I_j} c_{ij} \Delta_{ij} e^{-i(E^{(S)}_j - \Delta_{ij})t}\ket{E^{(S)}_j} = \bra{t}\sum_{j} \sum_{i\in I_j} c_{ij} \Delta_{ij} \ket{E^{(C)}_i} \ket{E^{(S)}_j}
\end{equation}
that is an identity, considering that in the right-hand side of the equation (\ref{44}) $\braket{t|E^{(C)}_i} = e^{iE^{(C)}_i t}$ and the global constraint on energy gives $E^{(C)}_i = - E^{(S)}_j + \Delta_{ij}$ (remember that the total energy term $E$ has been removed).

To verify that the (\ref{soluzione}) is a solution for (\ref{eqschrodinger2}) we have first to see how $\hat{\Delta}(t,t')$ acts on $\ket{\phi (t')}_S$. We have

\begin{equation}\label{41}
\hat{\Delta}(t,t') \ket{\phi (t')}_S = \sum_{j}\sum_{k \in I_j} \sum_{i \in I_j} e^{iE^{(C)}_k t} \Delta_{kj} e^{-iE^{(C)}_k t'}c_{ij} e^{iE^{(C)}_it'}\ket{E^{(S)}_j}.  
\end{equation}
So, by substituting the state (\ref{soluzione}) in equation (\ref{eqschrodinger2}) and using (\ref{41}), we obtain:

\begin{equation}
\sum_{j}\sum_{i\in I_j} c_{ij} \Delta_{ij} e^{-i(E^{(S)}_j - \Delta_{ij})t}\ket{E^{(S)}_j} = \frac{1}{T} \int_{t_0}^{t_0 + T} dt' \sum_{j}\sum_{k\in I_j} \sum_{i\in I_j} e^{iE^{(C)}_k t} \Delta_{kj} e^{-iE^{(C)}_k t'} c_{ij} e^{iE^{(C)}_it'}\ket{E^{(S)}_j}
\end{equation}
\begin{equation}
\Rightarrow \sum_{j}\sum_{i\in I_j} c_{ij} \Delta_{ij} e^{-i(E^{(S)}_j - \Delta_{ij})t}\ket{E^{(S)}_j} = \sum_{j}\sum_{i\in I_j} c_{ij} \Delta_{ij} e^{-i(E^{(S)}_j - \Delta_{ij})t}\ket{E^{(S)}_j}
\end{equation}
\end{widetext}
where in the last step we used again the constraint on the energy $E^{(C)}_i = - E^{(S)}_j + \Delta_{ij}$ and (see Appendix C for the proof):

\begin{equation}\label{delta}
\int_{t_0}^{t_0+T} dt'  e^{i (E^{(C)}_i - E^{(C)}_k)t'} =T \delta_{i,k} .
\end{equation}

\section{Proof of equation (\ref{delta})}

We start here considering a generic state $\ket{\psi} \in \mathcal{H}_C$, that is 
\begin{equation}\label{ultima}
\ket{\psi}= \sum_{k=0}^{p} c_k\ket{E^{(C)}_k} 
\end{equation}
where, we recall, $p+1=d_C$. 
We now apply in sequence the resolutions of the identity $\mathbb{1}_{C} = \frac{1}{T} \int_{t_0}^{t_0+T} dt \ket{t} \bra{t}$ and $\mathbb{1}_{C} = \sum_{n=0}^{p} \ket{E^{(C)}_n}\bra{E^{(C)}_n}$ thus obtaining:

\begin{equation}\label{hhh}
\begin{split}
\ket{\psi} & = \frac{1}{T} \int_{t_0}^{t_0+T} dt \ket{t}\braket{t|\psi} = \frac{1}{T} \int_{t_0}^{t_0+T} dt\sum_{k=0}^{p} c_k e^{it E^{(C)}_k} \ket{t}=\\ \\&
= \sum_{n=0}^{p}\ket{E^{(C)}_n}\bra{E^{(C)}_n}\frac{1}{T} \int_{t_0}^{t_0+T} dt\sum_{k=0}^{p} c_k e^{it E^{(C)}_k} \ket{t} =\\ \\&= \sum_{n=0}^{p} \sum_{k=0}^{p} c_k \frac{1}{T} \int_{t_0}^{t_0+T} dt e^{it (E^{(C)}_k - E^{(C)}_n)}\ket{E^{(C)}_n} .
\end{split}
\end{equation}
Since the state $\ket{\Psi}$ in equation (\ref{hhh}) has to be equal to (\ref{ultima}), 
we have $\int_{t_0}^{t_0+T} dt e^{it (E^{(C)}_k - E^{(C)}_n)}= T \delta_{k,n}$. 

\begin{widetext}
\section{Proof of equation (\ref{mediatemporale})}
	
Here we prove equation (\ref{mediatemporale}), namely we show that, although the states $\ket{t}$ are not orthogonal, we have

\begin{equation}\label{41b}
\hat{\rho}_S =Tr_C\left[\ket{\Psi}\bra{\Psi}\right]= \frac{1}{T} \int_{t_0}^{t_0+T} dt \braket{t|\Psi}\braket{\Psi|t}.
\end{equation}

Then, thanks to Canonical Typicality, we know that $\hat{\rho}_S \approx \frac{1}{Z} e^{-\beta \hat{H}_S}$ where $\beta = dS(E)/dE$ is the inverse temperature (with $S(E)$ the entropy of $C$) and $Z = Tr \left[ e^{-\beta \hat{H}_S}\right]$ \cite{canonical}. We start calculating the partial trace of the global state $Tr_C\left[\ket{\Psi}\bra{\Psi}\right]$ through the energy basis in the subspace $C$:

\begin{equation}\label{51}
\begin{split}
\hat{\rho}_S &= Tr_C\left[ \ket{\Psi}\bra{\Psi}\right] = \sum_{n} \braket{E^{(C)}_n|\Psi}\braket{\Psi|E^{(C)}_n} = \\&
=\sum_{n}  \sum_{j}\sum_{i\in I_j}  \sum_{l} \sum_{k \in I_l} c_{ij}c^{*}_{kl} \braket{E^{(C)}_n|E^{(C)}_i}\braket{E^{(C)}_k|E^{(C)}_n}\ket{E^{(S)}_j}\bra{E^{(S)}_l}
\end{split}
\end{equation}  
where $I_j$ is the set of the environment levels such that $E^{(C)}_i \in \left[E- E^{(S)}_j , E- E^{(S)}_j + \delta \right]$. Now, being $\delta  \ll \Delta E^{(S)}_j ~ \forall j$, then the energy intervals for the clock $\equiv$ environment coupled with each level $E^{(S)}_j$ of $S$ are pairwise disjoint. So equation (\ref{51}) becomes

\begin{equation}\label{asterisco}
\hat{\rho}_S = Tr_C\left[ \ket{\Psi}\bra{\Psi}\right] 
=  \sum_{j} \sum_{i\in I_j}\left|c_{ij} \right|^2 \ket{E^{(S)}_j}\bra{E^{(S)}_j} .
\end{equation}
Going instead to calculate the right-hand side of equation (\ref{41b}) we have
\begin{equation}
\begin{split}
\frac{1}{T} \int_{t_0}^{t_0+T} dt \braket{t|\Psi}\braket{\Psi|t} &= \frac{1}{T} \int_{t_0}^{t_0+T} dt \sum_{j}\sum_{i\in I_j} \sum_{l}  \sum_{k \in I_l} c_{ij}c^{*}_{kl} \braket{t|E^{(C)}_i}\braket{E^{(C)}_k|t} \ket{E^{(S)}_j}\bra{E^{(S)}_l}= \\&
= \sum_{j}\sum_{i\in I_j} \sum_{l}  \sum_{k \in I_l} c_{ij}c^{*}_{kl} \frac{1}{T} \int_{t_0}^{t_0+T} dt e^{-it(E^{(C)}_i - E^{(C)}_k)} \ket{E^{(S)}_j}\bra{E^{(S)}_l}
\end{split}
\end{equation}
and, considering $\int_{t_0}^{t_0+T} dt e^{-it(E^{(C)}_i - E^{(C)}_k)} = T\delta_{i,k}$ (see Appendix C), we obtain

\begin{equation}\label{54}
\frac{1}{T} \int_{t_0}^{t_0+T} dt \braket{t|\Psi}\braket{\Psi|t} 
= \sum_{j} \sum_{i\in I_j}\left|c_{ij} \right|^2 \ket{E^{(S)}_j}\bra{E^{(S)}_j}
\end{equation}
where we used again the fact that the relevant energy intervals for the clock $\equiv$ environment coupled with each level $E^{(S)}_j$ of $S$ are pairwise disjoint. Equation (\ref{54}) that is the same of (\ref{asterisco}), so $\frac{1}{T} \int_{t_0}^{t_0+T} dt \braket{t|\Psi}\braket{\Psi|t} = Tr_C\left[\ket{\Psi}\bra{\Psi}\right] = \hat{\rho}_S$.

\section{Proof of equation (\ref{evestgrande})}

We calculate here the expectation value $\langle \hat{X} \rangle_{t} = \bra{\phi(t)} \hat{X} \ket{\phi(t)}$, where $\hat{X}=\sqrt{\frac{1}{2m\omega}} (\hat{a} + \hat{a}^{\dagger})$, considering the relative state $\ket{\phi(t)}_S$ written as 

\begin{equation}
\ket{\phi(t)}_S = \sum_{i \in I_0} c_{i0} e^{-i (E^{(S)}_0 - \Delta_{i0})t} \ket{0^{(S)}} + \sum_{i \in I_1} c_{i1} e^{-i (E^{(S)}_1 - \Delta_{i1})t} \ket{1^{(S)}}.
\end{equation}
We have:

\begin{multline}
\langle \hat{X} \rangle_{t} = \left( \sum_{i \in I_0} c^{*}_{i0}e^{i(E^{(S)}_0 - \Delta_{i0})t}\bra{0^{(S)}} + \sum_{i \in I_1} c^{*}_{i1}e^{i(E^{(S)}_1 - \Delta_{i1})t}\bra{1^{(S)}} \right) \times \\
\times \sqrt{\frac{1}{2m\omega}} (\hat{a} + \hat{a}^{\dagger}) \left(  \sum_{i \in I_0} c_{i0}e^{-i(E^{(S)}_0 - \Delta_{i0})t}\ket{0^{(S)}} + \sum_{i \in I_1} c_{i1}e^{-i(E^{(S)}_1 - \Delta_{i1})t}\ket{1^{(S)}} \right)
\end{multline}
\begin{multline}\label{contox1}
\Rightarrow  \langle \hat{X} \rangle_{t} = \sqrt{\frac{1}{2m\omega}} \left( \sum_{i \in I_0} c^{*}_{i0}e^{i(E^{(S)}_0 - \Delta_{i0})t}\bra{0^{(S)}} + \sum_{i \in I_1} c^{*}_{i1}e^{i(E^{(S)}_1 - \Delta_{i1})t}\bra{1^{(S)}} \right) \times \\
\times \left( \sum_{i \in I_1}c_{i1} e^{-i(E^{(S)}_1 - \Delta_{i1})t}\ket{0^{(S)}} + \sum_{i \in I_0}c_{i0} e^{-i(E^{(S)}_0 - \Delta_{i0})t}\ket{1^{(S)}} + \sqrt{2} \sum_{i \in I_1}c_{i1} e^{-i(E^{(S)}_1 - \Delta_{i1})t}\ket{2^{(S)}}\right) .
\end{multline}
From equation (\ref{contox1}) we have:

\begin{multline}\label{fff}
\langle \hat{X} \rangle_{t} = \sqrt{\frac{1}{2m\omega}}  \sum_{i \in I_0} \sum_{k \in I_1} c^{*}_{i0} c_{k1} e^{i(E^{(S)}_0 - E^{(S)}_1 -\Delta_{i0} +\Delta_{k1})t } + \sqrt{\frac{1}{2m\omega}}  \sum_{i \in I_0} \sum_{k \in I_1} c_{i0} c^{*}_{k1} e^{-i(E^{(S)}_0 - E^{(S)}_1 -\Delta_{i0} +\Delta_{k1})t } .
\end{multline}
Now, writing $c_{i0}= |c_{i0}|e^{i\varphi_{i0}}$, $c_{k1}= |c_{k1}|e^{i\varphi_{k1}}$, $\Delta\varphi^{(0,1)}_{ik} = \varphi_{k1} - \varphi_{i0}$ and considering that $E^{(S)}_1 - E^{(S)}_0 = \omega$, we obtain:

\begin{equation}\label{contox2}
\langle \hat{X} \rangle_{t} = \sqrt{\frac{2}{m\omega}} \sum_{i \in I_0} \sum_{k \in I_1}  |c_{i0}||c_{k1}|\cos((\omega  + (\Delta_{i0} - \Delta_{k1}))t - \Delta\varphi^{(0,1)}_{ik}) 
\end{equation}
which is what we had to prove, since equation (\ref{contox2}) is the same of the second part of equation (\ref{evestgrande}). 

If we want instead to consider the expectation value $\langle \hat{X} \rangle_{t}$ expressed in terms of the $\alpha_j(t)$, through the definition (\ref{alphat}) (that is $\alpha_j (t) = \sum_{i \in I_j} c_{ij}e^{i \Delta_{ij} t}$), we can rewrite equation (\ref{fff}) as

\begin{equation}\label{ffff}
\langle \hat{X} \rangle_{t} = \sqrt{\frac{1}{2m\omega}} \left(  \alpha^{*}_{0}(t) \alpha_{1}(t) e^{i(E^{(S)}_0 - E^{(S)}_1)t } +  \alpha_{0}(t) \alpha^{*}_{1}(t) e^{-i(E^{(S)}_0 - E^{(S)}_1)t } \right).
\end{equation}
Writing $\alpha_{0}(t)= |\alpha_{0}(t)|e^{i\phi_{0}(t)}$, $\alpha_{1}(t)= |\alpha_{1}(t)|e^{i\phi_{1}(t)}$, $\Delta\phi (t)= \phi_{1}(t) - \phi_{0}(t)$ and considering again $E^{(S)}_1 - E^{(S)}_0 = \omega$, (\ref{ffff}) becomes:

\begin{equation}\label{ggg}
\langle \hat{X} \rangle_{t} =\sqrt{\frac{2}{m\omega}} |\alpha_{0}(t)||\alpha_{1}(t)|\cos(\omega t - \Delta\phi(t)).
\end{equation}
Equation (\ref{ggg}) is the same of the first part of equation (\ref{evestgrande}) and shows the expectation value $\langle \hat{X} \rangle_{t}$ expressed in terms of the time-dependent coefficients $\alpha_j(t)$.

\section{Proof of equation (\ref{dimappe})}

We prove here that equation (\ref{evestgrande}) reduces to (\ref{dimappe}) in the case of $t \ll 1/|\Delta_{i0} - \Delta_{k1}|$. We start considering the second part of equation (\ref{evestgrande}), that is: 	
\begin{equation}\label{evestgrandeapp}
\begin{split}
\langle \hat{X} \rangle_{t} 
= \sqrt{\frac{2}{m\omega}} \sum_{i \in I_0} \sum_{k \in I_1}  |c_{i0}||c_{k1}|\cos(\omega t  + (\Delta_{i0} - \Delta_{k1})t - \Delta\varphi^{(0,1)}_{ik}) .
\end{split}
\end{equation}
This can be rewritten as

\begin{equation}
\begin{split}
\langle \hat{X} \rangle_{t} &= \sqrt{\frac{2}{m\omega}} \left( \sum_{i \in I_0} \sum_{k \in I_1}|c_{i0}||c_{k1}|\cos(\omega t  - \Delta\varphi^{(0,1)}_{ik})\cos( (\Delta_{i0} - \Delta_{k1})t) \right) + \\& 
-       \sqrt{\frac{2}{m\omega}}  \left( \sum_{i \in I_0} \sum_{k \in I_1}|c_{i0}||c_{k1}| \sin(\omega t  - \Delta\varphi^{(0,1)}_{ik})\sin((\Delta_{i0} - \Delta_{k1})t)  \right) .
\end{split}
\end{equation}
We impose now the condition $t \ll 1/|\Delta_{i0} - \Delta_{k1}|$, so we can consider the Taylor expansions of $\cos((\Delta_{i0} - \Delta_{k1})t)$ and $\sin((\Delta_{i0} - \Delta_{k1})t)$ thus obtaining

\begin{equation}\label{hp}
\begin{split}
\langle \hat{X} \rangle_{t} &= \sqrt{\frac{2}{m\omega}} \sum_{i \in I_0}\sum_{k \in I_1}|c_{i0}||c_{k1}|\cos(\omega t - \Delta\varphi^{(0,1)}_{ik}) + \\& - \sqrt{\frac{2}{m\omega}} \sum_{i \in I_0}\sum_{k \in I_1}|c_{i0}||c_{k1}| t(\Delta_{i0} - \Delta_{k1})\sin(\omega t  - \Delta\varphi^{(0,1)}_{ik}) +  \\&
- \sqrt{\frac{1}{2m\omega}}   \sum_{i \in I_0}\sum_{k \in I_1}|c_{i0}||c_{k1}|t^2(\Delta_{i0} - \Delta_{k1})^2\cos(\omega t  - \Delta\varphi^{(0,1)}_{ik}) + ... \:\: . 
\end{split}
\end{equation}
With $\alpha_0(0) = |\alpha_0(0)|e^{i \phi_0(0)}$, $\alpha_1(0) = |\alpha_1(0)|e^{i \phi_1(0)}$, $\Delta \phi(0) = \phi_1(0) - \phi_0(0)$ and $\alpha_j(0) = \sum_{i \in I_j} c_{ij}$, we have 

\begin{equation}
\sum_{i \in I_0} \sum_{k \in I_1}  |c_{i0}||c_{k1}|\cos(\omega t  - \Delta\varphi^{(0,1)}_{ik}) =  |\alpha_0(0)||\alpha_1(0)|\cos(\omega t - \Delta \phi(0)) .
\end{equation}
\end{widetext}
So, considering up to the first order of approximation in $t( \Delta_{i0} - \Delta_{k1} )$, we have finally for the expectation value of the position operator $\langle \hat{X} \rangle_{t}$:

\begin{multline}\label{58}
\langle \hat{X} \rangle_{t} \simeq  \sqrt{\frac{2}{m\omega}} |\alpha_0(0)||\alpha_1(0)|\cos(\omega t - \Delta \phi(0)) +\\-  \sqrt{\frac{2}{m\omega}}\sum_{i \in I_0} \sum_{k \in I_1}  |c_{i0}||c_{k1}| t( \Delta_{i0} - \Delta_{k1} )\sin(\omega t  - \Delta\varphi^{(0,1)}_{ik}) 
\end{multline}
that is what we needed to show being (\ref{58}) the same of equation (\ref{dimappe}).


	
\end{document}